\documentclass[12pt]{article}

\newcommand{\bqn}{\begin{eqnarray}}
\newcommand{\eqn}{\end{eqnarray}}
\newcommand{\be}{\begin{equation}}
\newcommand{\ee}{\end{equation}}

\begin{document}

\title{The One Dimensional Damped Forced Harmonic Oscillator Revisited} 
\author{G. Flores-Hidalgo, F.A. Barone\\
{\it ICE - Universidade Federal de Itajub\'a, Av. BPS 1303,}\\
{\it Caixa Postal 50 - 37500-903, Itajub\'a, MG, Brazil.}} 

\maketitle

\begin{abstract}
In this paper we give a general solution to the problem of the damped harmonic oscillator under the influence of an arbitrary time-dependent external force. We employ simple methods accessible for beginners and useful for undergraduate students and professors in an introductory course of mechanics.
%We employ simple methods accessible for beginners at university level and useful for undergraduate students and professors in an introductory course of mechanics. 
\end{abstract}

One of the most important, and investigated, problems in physics is the harmonic oscillator. This fact is due to the relevance of this problem in the classical as well as in the quantum contexts and, also, because it is one of the few problems we know how to solve exactly. The solution to this problem is given in many textbooks of classical mechanics, both at introductory and advanced levels, and is given with two arbitrary constants determined according to the initial conditions. 
The treatment of the damped harmonic oscillator with the presence of external forces can also be found in the literature and its formal treatment, for any time dependent external force, requires the formalism of Green's functions. Generally, undergraduate students at introductory levels are not yet acquainted with Green's functions formalism, so this subject is usually omitted at a first course of mechanics where it is considered only the case of an external force with sinus-like dependence. In this last case it is interesting to call attention to the treatment given in Ref. \cite{weinstock} where the problem is solved without resorting to the standard theory of second order differential equation. Also, a similar treatment is given in Ref. \cite{bush} for the single harmonic oscillator. On the other hand, there are also elaborated treatments for this problem, particularly interesting is the Feynman diagrams technique given in Ref. \cite{thorndike} and the Poisson bracket formalism presented in Ref. \cite{farina}.

In this work, which is intended for undergraduate students and professors, we give a general solution to the problem of the damped harmonic oscillator under the influence of an arbitrary time-dependent external force. The main advantages of the procedure proposed in this paper in comparison with the ones encountered in standard textbooks are: the constants are fixed from the beginning, according to the initial conditions of the problem, the mathematical methods employed are accessible for beginners once they are not sophisticated, we can consider an external force with any time dependence and not only the sinus-like one and the method resembles the factorization employed for solving the quantum harmonic oscillator. 

The differential equation which governs the time evolution of the damped harmonic oscillator with external time dependent force is
%%%%%%%%%
\begin{equation}
m\frac{d^2x(t)}{dt^2}=-b\frac{dx(t)}{dt}-kx(t)+F(t)\;,
\label{newton}
\end{equation}
%%%%%%%%%
where $b$ is the damping coefficient, $k$ is the restoring constant force, $m$ is the mass of the oscillator and $F(t)$ is the external force. The dynamics of the oscillator is determined by solving equation (\ref{newton}) with the initial conditions, that is, the initial position and velocity, $x(t_0)=x_{0}$ and $\dot{x}(t_0)=v_{0}$, respectively.

In order to solve for $x(t)$ we first define the constants $2\gamma=b/m$ and $\omega_0^2=k/m$ and the differential operator $D=d/dt$. In this way we write Eq. (\ref{newton}) in the form
%%%%%%%%%
\begin{equation}
\left(D^2+2\gamma D+\omega_0^2\right)x(t)=F(t)\;,
\label{operator}
\end{equation}
%%%%%%%%%
which is equivalent to
%%%%%%%%%
\begin{equation}
\label{prod}
(D-\alpha)(D-\beta)x(t)=F(t)\;,
\end{equation}
%%%%%%%%%
where $\alpha$ and $\beta$ are the roots of the quadratic equation $y^2-2\gamma y +\omega_0^2=0$, that is, $\alpha=-\gamma+i\omega$, $\beta=-\gamma-i\omega$, where $\omega=\sqrt{\omega_0^2-\gamma^2}$.

Using the fact that
%%%%%%%%%
\begin{equation}
\label{zxc1}
(D-a)f(t)=e^{at}D(e^{-at}f(t))\ ,
\end{equation}
%%%%%%%%%
which is valid for any complex constant $a$ and any differentiable function $f(t)$, we can rewrite Eq. (\ref{prod}) as follows
%%%%%%%%%
\begin{equation}
\label{zxc2}
(D-\alpha)\Bigl[e^{\beta t}D\Bigl(e^{-\beta t}x(t)\Bigr)\Bigr]=F(t)\ ,
\end{equation}
%%%%%%%%%
where we have made $a=\beta$ and $f(t)=x(t)$.

Now we use property (\ref{zxc1}) again, but with $a=\alpha$ and $f(t)=e^{\beta t}D\Bigl(e^{-\beta t}x(t)\Bigr)$, so Eq. (\ref{zxc2}) becomes
%%%%%%%%%
\begin{equation}
e^{\alpha t}D\Bigl[e^{-\alpha t}\ e^{\beta t}D\Bigl(e^{-\beta t}x(t)\Bigr)\Bigr]=F(t)\ \Rightarrow\ D\Bigl[e^{-(\alpha-\beta) t}D\Bigl(e^{-\beta t}x(t)\Bigr)\Bigr]=e^{-\alpha t}F(t).
\end{equation}
%%%%%%%%%

Integrating in the time variable and using the initial conditions we have
%%%%%%%%%
\begin{equation}
\label{casi}
e^{-(\alpha-\beta) t}D\Bigl(e^{-\beta t}x(t)\Bigr)-e^{-(\alpha-\beta) t_0}\Bigl(-\beta e^{-\beta t_0}x_{0}+e^{-\beta t_0}v_{0}\Bigr)=\int_{t_0}^{t} e^{-\alpha t'}F(t')dt'\;,
\end{equation}
%%%%%%%%%
where we have performed some simple manipulations on the second term on the left hand 
side of the above equation.

From Eq. (\ref{casi}) we have
%%%%%%%%%
\begin{equation}
D\Bigl(e^{-\beta t}x(t)\Bigr)=e^{(\alpha-\beta)t}\Biggl[e^{-\alpha t_0}\Bigl(-\beta x_{0}+v_{0}\Bigr)+\int_{t_0}^{t} e^{-\alpha t'}F(t')dt'\Biggr]\ .
\end{equation}
%%%%%%%%%

Performing a second integration and arranging terms we obtain,
%%%%%%%%%
\begin{equation}
\label{pen}
x(t)=x_{0}e^{\beta(t-t_{0})}+\frac{v_{0}-\beta x_{0}}{\alpha-\beta}\Bigl(e^{\alpha (t-t_0)}-e^{\beta (t-t_0)}\Bigr)
+e^{\beta t}\int_{t_0}^{t}\int_{t_0}^{t''}e^{-\alpha t'}F(t')dt'\ e^{(\alpha-\beta) t''}dt''\ .
\end{equation}
%%%%%%%%%

In order to have just one integration involving the external force $F(t)$, let us manipulate the last term of Eq. (\ref{pen}). By using the standard expression for integration by parts
%%%%%%%%%
\begin{equation}
\int_{t_{0}}^{t}U(t'')dV(t'')=U(t)V(t)-U(t_{0})V(t_{0})-\int_{t_{0}}^{t}V(t'')dU(t'')\ ,
\end{equation}
%%%%%%%%% 
with
%%%%%%%%%
\begin{eqnarray}
U(t'')&=&\int_{t_{0}}^{t''}e^{-\alpha t'}F(t')dt'\ \Rightarrow\ dU(t'')=e^{-\alpha t''}F(t'')dt''\ ,\cr\cr
dV(t'')&=&e^{(\alpha-\beta)t''}dt''\ \Rightarrow\  V(t'')=\frac{1}{\alpha-\beta}\ e^{(\alpha-\beta)t''}\ ,
\end{eqnarray}
%%%%%%%%%
we can write
%%%%%%%%%
\begin{equation}
\label{intpartes}
\int_{t_0}^{t}\int_{t_0}^{t''}e^{-\alpha t'}F(t')dt'\ e^{(\alpha-\beta) t''}dt''=\frac{e^{(\alpha-\beta)t}}{\alpha-\beta}
\int_{t_{0}}^{t}e^{-\alpha t'}F(t')dt'-\frac{1}{\alpha-\beta}\int_{t_{0}}^{t}e^{-\beta t'}F(t')dt'\ .
\end{equation}
%%%%%%%%%

Substituting (\ref{intpartes}) into Eq. (\ref{pen}), using the definitions of $\alpha$ and $\beta$ given after Eq. (\ref{prod}) and performing some simple manipulations we have finally
%%%%%%%%%
\begin{eqnarray}
x(t)&=&\frac{e^{-\gamma(t-t_0)}}{\omega}\Biggl[x_{0}\left[\omega\cos\Bigl(\omega(t-t_0)\Bigr)+
\gamma \sin\Bigl(\omega(t-t_0)\Bigr)\right]+v_{0}\sin\Bigl(\omega(t-t_0)\Bigr)\Biggr]\cr\cr
&\ &+\frac{e^{-\gamma t}}{2i\omega}\left[e^{i\omega t}\int_{t_0}^{t}e^{(\gamma-i\omega)t'}
F(t')dt'-e^{-i\omega t}\int_{t_0}^{t}e^{(\gamma+i\omega)t'}F(t')dt'\right]\ .
\label{solutiona}
\end{eqnarray}
%%%%%%%%%
Above expression can be written in a more familiar way
\begin{eqnarray}
x(t)&=&\frac{e^{-\gamma(t-t_0)}}{\omega}\Biggl[x_{0}\left[\omega\cos\Bigl(\omega(t-t_0)\Bigr)+
\gamma \sin\Bigl(\omega(t-t_0)\Bigr)\right]+v_{0}\sin\Bigl(\omega(t-t_0)\Bigr)\Biggr]\cr\cr
&\ &+\frac{1}{\omega}\int_{t_0}^t e^{-\gamma(t-t')}\sin\Bigl(\omega(t-t')\Bigr)
F(t')dt'\;,
\label{solutionb}
\end{eqnarray}
in complete agreement with the one encountered in standard textbooks, see for example Ref. \cite{symon}.

%%%%%%%%%

%%%%%%%%%


\begin{thebibliography}{99}

\bibitem{weinstock} R. Weinstock, \lq\lq An Unusual Method for Solving the Harmonic-Oscillator Equation\rq\rq, Am. J. Phys. {\bf 29}, 830-831 (1961).

\bibitem{bush} R.T. Bush, \lq\lq The simple harmonic oscillator: an alternative solution of the equation for damped oscillation\rq\rq, Am. J. Phys. {\bf 41}, 738-739 (1973).

\bibitem{thorndike} A. Thorndike, \lq\lq Using Feynman diagrams to solve the classical harmonic oscillator\rq\rq, Am. J. Phys. {\bf 68}, 155–159 (2000).

\bibitem{farina} C.F. Farina and M.M. Gandelman, \lq\lq An algebraic approach for solving mechanical problems\rq\rq, Am. J. Phys. {\bf 58}, 491-495 (1990).

\bibitem{symon} K.R. Symon, {\it Mechanics}, (Addison-Wesley Reading, MA, 1972), 3$^{rd}$ ed.
      
\end{thebibliography}
\end{document}